\documentclass[twocolumn, showpacs]{revtex4}

\usepackage{amsmath}
\usepackage{amssymb}
\usepackage{graphicx}
\usepackage{calrsfs}

\newcommand{\be}{\begin{equation}}
\newcommand{\ee}{\end{equation}}

\newcommand{\Arg}{\mathrm{Arg}}
\newcommand{\Ln}{\mathrm{Ln}}

\newcommand{\rmm}{\mathrm{m}}
\newcommand{\rmTE}{\mathrm{TE}}
\newcommand{\rmTM}{\mathrm{TM}}

\newcommand{\vkp}{\mathbf{k}_\perp}

\newcommand{\kp}{k_\perp}
\newcommand{\kB}{k_\mathrm{B}}
\newcommand{\im}{\Im\text{m}}

\newcommand{\infintw}{\int_{-\infty}^{\infty}\!\!\!d\omega}
\newcommand{\infintnw}{\int_{0}^{\infty}\!\!\!d \omega}
\newcommand{\infintk}{\int_0^\infty d \kp ~\kp}

\newcommand{\qsum}{\sum_{q=\rmTE}^\rmTM}

\newcommand{\op}{\omega_\mathrm{p}}  %plasma frequency

\begin{document}

\title{Nernst's heat theorem for Casimir-Lifshitz free energy}
\date{\today}
\author{Simen A. \surname{Ellingsen}}
\affiliation{Department of Energy and Process Engineering, Norwegian University of Science and Technology, N-7491 Trondheim, Norway}\email{simen.a.ellingsen@ntnu.no}

\begin{abstract}
  By regarding the Lifshitz expression for the Casimir free energy on the real frequency axis rather than the imaginary Matsubara frequencies as is customary, new light is shed on the ongoing debate regarding the thermodynamical consistency of this theory in combination with common permittivity models. It is argued that when permittivity is temperature independent over a temperature interval including zero temperature, a cavity made of causal material with continuous dispersion properties separated by vacuum cannot violate Nernst's theorem (the third law of thermodynamics). The purported violation of this theorem pertains to divergencies in the double limit in which frequency and temperature vanish simultaneously. While any model should abide by the laws of thermodynamics within its range of applicability, we emphasise that the Nernst heat theorem is a relevant criterion for choosing amongst candidate theories only when these theories are fully applicable at zero temperature and frequency.
\end{abstract}


\maketitle

Since Bostr\"{o}m and Sernelius first predicted the existence of large thermal corrections to the Casimir force in 2000\cite{bostrom00}, controversies over the thermal behaviour of this effect, which in its most typical embodiment may be seen as the attraction between macroscopic objects due to zero-point fluctuations of the quantum vacuum, have been extensively covered in the published literature. The use of the Drude model to describe dielectric permittivity, employed in \cite{bostrom00} was soon criticised on thermodynamical grounds \cite{klimchitskaya01}. The reason was that in the case of a perfect crystal lattice, when all dissipation is due to scattering of electrons on thermal phonons, the Casimir free energy as calculated with the Lifshitz formula appears to violate the Nernst's heat theorem which states that entropy should vanish as $T\to 0$. The Drude model was defended by other authors \cite{hoye03,bostrom04,brevik05} who argued that since the Drude model offers better description of impure metals, and since real metal samples always have impurities, the Drude model must be employed. It was shown in 2003 \cite{brevik04} and recently in a more extensive treatment\cite{hoye07} that the free energy with Drude permittivity is quadratic in $T$ for small temperatures when impurities are present. No consensus has yet been reached on the important physical question of why Casimir force predictions for the perfect lattice model, important in solid state physics, differs significantly from those pertaining to real metals with a very small but nonzero concentration of imperfections.

Recently, a somewhat analoguous problem was brought forth for dielectrics with a small conductivity for finite $T$ which vanishes at $T=0$\cite{geyer05}. While the purported violation in the case of Drude metals referred to the transverse electric (TE) mode, this time the bother appears to be a discontinuity in the transverse magnetic (TM) Fresnel reflection coefficient giving rise to nonzero entropy at zero temperature. The problem was recently argued to extend to insulators, intrinsic and lightly doped semiconductors as well as Mott-Hubbard semiconductors and indeed the permittivity contribution from Debye rotation of molecular dipoles\cite{geyer08}.

According to the argument presented herein both the Drude model and the dielectric permittivity model with conductivity included belong to a group of permittivities which cannot violate Nernst's theorem when permittivity can be regarded as temperature invariant in a range of temperatures near and including $T=0$. 

While the present paper does not aspire to solve the physical question of how to take into account the presence of a small conductivity in dielectrics when substituted into the Lifshitz formula, it seeks to illuminate the ever recurring question of thermodynamical consistency. It has previously been shown\cite{milton04} by use of the Euler-Maclaurin (or equivalently Abel-Plana) formula that for Fresnel reflection coefficients which are continuous functions of \emph{imaginary} frequencies in the limit $T\to 0$, Nernst's theorem is satisfied. An exploration of Casimir entropy in the formalism of surface modes was undertaken independently of this work by Intravaia and Henkel \cite{intravaia07} whose conclusions accord with ours. By a method of summation of the eigenmodes of the vacuum between two plates they demonstrate that Nernst's theorem can be broken between metal plates only for temperature dependent relaxation such as in a perfect and infinitely large metal lattice. 

This paper demonstrates a similar result using the real frequency Lifshitz formalism between plates of a generic nonmagnetic materials whose permittivity satisfies a small set of criteria. The real frequency formalism is more complicated and less elegant, but with the advantage of a more direct physical interpretation. Finally, a discussion of the physical implications of the mathematical limits involved is given. In particular we emphasise the importance of assessing when Nernst's theorem, which concerns zero temperature, can be used to inform finite temperature physics. 

\section{Introduction: Free energy and entropy at real frequencies}

The Lifshitz expression \cite{lifshitz56} for the free energy per unit transverse area between two identical dielectric plates separated by vacuum is in general of the form %(obtained for example through the stress tensor derivation of the pressure \cite{schwinger78} and integration with respect to $a$)
\be \label{Fim}
  \mathcal{F}(a,T) = \infintnw \coth\left( \frac{\omega}{2\omega_T}\right)\im\{\phi(\omega, T)\},
\ee
where $\omega_T\equiv\kB T/\hbar$ and $\phi(\omega,T)$ is the zero temperature integrand
\be\label{phi}
  \phi(\omega,T) = \frac{\hbar}{4\pi^2}\infintk\qsum \ln D_q(\omega,c\vkp,T)
\ee
wherein
\begin{subequations}
  \begin{align}\label{Dq}
    D_q &\equiv 1-r_q^2 \exp(-2\kappa_0 a); \\
    r_\rmTE &= \frac{\kappa_0-\kappa}{\kappa_0+\kappa}; ~~~r_\rmTM = \frac{\epsilon\kappa_0-\kappa}{\epsilon\kappa_0+\kappa}. 
  \end{align}
\end{subequations}
We have assumed $\mu=1$ everywhere for simplicity, $\epsilon$ is the permittivity of the dielectric relative to vacuum and
\be
  \kappa_0 = (\kp^2 - \omega^2/c^2)^{1/2}; ~~~\kappa = (\kp^2 -\epsilon\omega^2/c^2)^{1/2}.
\ee
We assume that $\epsilon$ does not depend on transverse momentum, thus neglecting any nonlocal effects and furthermore that it is a generalised susceptibility and obeys causality, which implies in particular that \cite{landau80}
\begin{enumerate}
  \item $\epsilon(-\omega^\ast) = \epsilon^\ast(\omega)$\label{epsilonSymm}
  \item $|\Im\mathrm{m}\{\epsilon(\omega)\}|>0$ on the entire real frequency axis except at $\omega=0$ where it may be undefined. \label{epsilonImag}
\end{enumerate}
In general $\epsilon$ is also temperature dependent, making for the temperature dependence of $\phi(\omega,T)$. The complex conjugate is denoted with an asterisk and we will consider only real frequencies henceforth. One might furthermore impose the physically reasonable demand that
\begin{enumerate}\setcounter{enumi}{2}
  \item $\epsilon(\omega)$ is continuous and $|\epsilon(\omega)|<\infty$ for all real frequencies except possibly $\omega=0$.\label{epsilonCont}
\end{enumerate}
%In particular, criterion \ref{epsilonCont} is fulfilled if $\epsilon(\omega)$ is continuous for all $\omega$ and obey criterion \ref{epsilonImag}.

%Examples of permittivities satisfying \ref{epsilonSymm} and \ref{epsilonImag} are 
%\begin{align}
%  \epsilon(\omega) &= 1 - \frac{\op}{\omega(\omega+ i \nu)} &\text{(Drude model)}\label{Drude}\\
%  \epsilon(\omega) &= 1 + \frac{\epsilon_0-1}{1- \omega^2/\omega_0^2} +  i  \frac{\sigma}{\omega} &\text{(Semiconductor)}\label{epDC}
%\end{align}
%with $\op$ the plasma frequency, $\nu$ the relaxation frequency and $\epsilon_0$ and $\omega_0$ material parameters. Some common models which do not obey \ref{epsilonImag} are
%\begin{align}
%  \epsilon(\omega) &= 1 - \frac{\op}{\omega^2} &\text{(Plasma model)}\label{epPlasma}\\
%  \epsilon(\omega) &= 1 + \frac{\epsilon_0-1}{1- \omega^2/\omega_0^2}  &\text{(Dielectric)} \label{epNoDC}
%\end{align}
%Evidently, models which do not obey criteria \ref{epsilonSymm}-\ref{epsilonCont} may still satisfy Nernst's theorem; such are not considered here.

%Note that (\ref{epDC}) and (\ref{epNoDC}) have simple poles at $\omega=\pm\omega_0$, causing $r_q$ to jump from $-1$ to $1$ or vice versa while the \emph{squared} reflection coefficient is continuous in keeping with criterion \ref{epsilonCont}. In a more realistic model, the denominators of the divergent fraction would contain an imaginary part as well, rendering the permittivity finite (the present form corresponds to a dissipation curve which is a $\delta$-function at frequency $\omega_0$ \cite{sernelius01}).

The function $\phi$ obeys the symmetry property $\phi(-\omega)=\phi^\ast(\omega)$ for real frequencies \footnote{This is easily argued: Because $\epsilon$ satisfies this relation by assumption, so, one finds, does $D_q$. The logarithm of a complex function is infinitely degenerate, and for $\ln D_q$ to give meaning we should interpret it as its principle value, $\Ln D_q \equiv \ln |D_q| + i\Arg D_q$, which incidentally also satisfies $\Ln D_q(-\omega)=[\Ln D_q(\omega)]^\ast$.}, hence the real part of $\phi(\omega)$ is even with respect to $\omega$ whilst its imaginary part is odd. This allows us to write $\mathcal{F}$ in a form which makes the mathematical discussion in the following somewhat more transparent. Since both $\im \phi$ and $\coth (\omega/2\omega_T)$ are odd functions of $\omega$, the integrand of (\ref{Fim}) is even and we can let the $\omega$ integral run from $-\infty$ to $\infty$ and divide by $2$. Adding the real part of $\phi$ by substituting $\im \phi \to \phi/i$ will make no difference, since it makes for an odd integrand term which vanishes under symmetrical integration, so
\be\label{F}
  \mathcal{F}(a,T) = \frac{1}{2i}\infintw \coth\left( \frac{\omega}{2\omega_T}\right)\phi(\omega, T)
\ee
is equivalent to (\ref{Fim}).

%, only the imaginary part of $\phi(\omega, T)$ with respect to $\omega$ contributes to (\ref{F}). %This is reflected in the original paper by Lifshitz\cite{lifshitz56}, where the integral is over positive $\omega$ only, but the imaginary part is taken\footnote{Allowing for the differences due to substitution to the variable $p$. See below.}.

Assume for the moment that $\epsilon$, and hence $\phi$, is invariant with temperature over at least a finite range of small temperatures including $T=0$. In this case the temperature dependence of $\mathcal{F}(T)$ can be treated very simply when $T$ is in this range, since the $T$ dependence now sits only in the factor $\coth(\omega/2\omega_T)$. %By writing
%\be \label{coth}
%  \coth\frac{\omega}{2\omega_T} = \sg (\omega)\left( 1 + \frac{2}{\exp(|\omega|/\omega_T)-1}\right),
%\ee
%$\mathcal{F}(T)$ now splits naturally into a $T$-independent part (the free energy at zero temperature) plus a temperature correction. $\sg$ is the signum function, so both terms of (\ref{coth}) are odd functions of $\omega$.

From thermodynamics the Casimir entropy in the cavity, $S$, is given as
\be
  S = -\frac{\partial}{\partial T} \mathcal{F}(T) ,
\ee
so if one were able to interchange integration and differentiation with respect to $T$, one could write
\begin{align}
    S &= -\frac{1}{2 i }\infintw \phi(\omega) \frac{d}{d T}\coth\left(\frac{\omega}{2\omega_T}\right)\notag\\
    &=-\frac{\hbar}{ i \kB T^2}\infintw \phi(\omega)\frac{\omega \exp(\omega/\omega_T)}{[\exp(\omega/\omega_T)-1]^2}.\label{integral2}
\end{align}
%We were able to skip the absolute value sign in the exponentials in the last line since the fraction is symmetrical under $\omega\to-\omega$ in exponentials. 

%A brief note on (\ref{coth}) may be required. Because the function $\sg(\omega)$ is not defined at $\omega=0$, it may not be obvious that the analytic properties of $\coth$ are preserved here. One easily verifies that it holds identically for all $\omega\neq 0$ including the limit $\omega\to 0$, and the expression (\ref{integral2}) could just as easily be found using the identity
%\be\label{coth2}
%  \coth\frac{\omega}{2\omega_T} = 1 + \frac{2}{\exp(\omega/\omega_T)-1}.
%\ee
%This expansion avoids this ambiguity\footnote{Indeed, (\ref{coth2}) follows directly from the definition of $\coth$.} but has the intuitive disadvantage that each term is not odd. Physically the single point $\omega=0$ cannot be of importance if the limit is unique (when $\epsilon$ varies with $T$ at $T=0$ we will see that this may no longer be the case) and since the original expression (\ref{Fim}) depends only on the \emph{limit} $\omega\to 0$, the single point $\omega=0$ can be disregarded.

For any finite $\omega$ the integrand of (\ref{integral2}) vanishes extremely fast, as $\exp(-\hbar|\omega|/\kB T)$, when $T\to 0$. This demonstrates the finding of Torgerson and Lamoreaux \cite{torgerson04} that temperature corrections are important only for frequencies below $\omega_T$, which is a very low frequency even at room temperature ($\sim 10^{13}$rad/s). At $\omega=0$ (and finite $T$) the rightmost fraction in (\ref{integral2}) has a simple pole, yet only the imaginary part of $\phi(\omega)$ contributes to (\ref{integral2}), which is zero here since $\Im\rmm\{\phi\}$ is an odd function of $\omega$, removing this pole. In the limit $T\to 0$, thus, entropy vanishes as it should and the third law of thermodynamics is obeyed (there are subtleties pertaining to the TM mode as will be discussed in the following).

Two questions arise from this consideration: Under which circumstances may differentiation be interchanged with integration? And what happens if one or more parameters of $\epsilon(\omega)$ are temperature dependent all the way down to zero temperature?

%%%%%%%%%%%%%%%%%%%%%%%%%%%%%%%%%%%%%%%%%
\section{Temperature independent $\epsilon(\omega)$}

We will treat the first question first. Leibnitz' integral rule for improper integrals,
\be \label{leibnitz}
  \frac{d}{d y}\int_{x_0}^{\infty}d x f(x,y) = \int_{x_0}^\infty d x \frac{d}{d y}f(x,y),
\ee
is always valid when  (\cite{whittaker62} \S4.44)
\begin{itemize}
  \item$f(x,y)$ and $d f(x,y)/d y$ are both continuous on $x\in [x_0,\infty\rangle$ and the relevant interval of values of $y$, 
  \item the integral on the left exists, and 
  \item the integral on the right converges uniformly.
  \end{itemize}
The generalisation to integrals with both limits infinite is trivial.

To make our considerations more concrete, let us concentrate on some permittivity models which are in common use:
\begin{align}
  \epsilon(\omega) &= 1 - \frac{\op^2}{\omega(\omega+ i \nu)} \label{Drude}\\
  \epsilon(\omega) &= 1 + \frac{\epsilon_\infty-1}{1- \omega^2/\omega_0^2 - i\gamma\omega/\omega_0^2} + \frac{ i  \sigma}{\epsilon_0\omega} \label{epDC}
\end{align}
of which the former is the Drude model for metals, and the latter describes a semiconductor. Here $\op$ is the plasma frequency, $\nu$ the relaxation frequency, $\epsilon_0$ the vacuum permittivity and $\epsilon_\infty , \gamma$ and $\omega_0$ material parameters. $\sigma$ is the DC conductivity of the semiconductor. Some common models which obey \ref{epsilonSymm} but not \ref{epsilonImag} are
\begin{align}
  \epsilon(\omega) &= 1 - \frac{\op^2}{\omega^2}    \label{epPlasma}\\
  \epsilon(\omega) &= 1 + \frac{\epsilon_\infty -1}{1- \omega^2/\omega_0^2},  \label{epNoDC}
\end{align}
the plasma model\footnote{A generalised, causal form of the plasma model was recently proposed \cite{geyer07}.} for metals and a model of dielectrics with $\delta$ function dissipation at $\omega=\omega_0$ \cite{sernelius01}. Notice that (\ref{Drude}) and (\ref{epDC}) obey criteria \ref{epsilonSymm}-\ref{epsilonCont}.

%%%%%%%%%%%%%%%%%%%%%%%%%%%%%%%%%%%%%
\subsection{Propagating and evanescent waves}

One notices that we seem to run into trouble with the continuity criterion at $|\omega| = c\kp$, where $\kappa_0=0$, since,  when regarded as a double integral, (\ref{F}) seems at first glance to imply integration \emph{across} the lines $\omega=\pm c\kp$, which would cause trouble with continuity: one sees from (\ref{Dq}) that $D_q=0$ for $\kappa_0=0$, hence the real part of $\Ln D_q$ is undefined and the imaginary part turns out to be discontinuous as these lines are crossed. 

The problem can be avoided, however. Let us define $\beta = \kp c/\omega$ for short. For positive frequencies, $\beta=1$ is the limit in which the electromagnetic fields in the cavity travel parallel to the plates and become evanescent in vacuum as the $\beta=1$ barrier is crossed, a limit whose discontinuous properties are physically obvious: the waves just on the propagating side ($c\kp$ just smaller than $\omega$) travel through the system just gracing the surfaces, while the fields on the evanescent side stay \emph{on} the surfaces; they are qualitatively different phenomena and the transition from one to the other can be expected to be discontinuous \footnote{Due to the nonzero imaginary part of $\epsilon$ there are no similar problems for $\kappa$ near $c\kp=|\omega|\sqrt{\Re\mathrm{e}\{\epsilon(\omega)\}}$.}. Negative frequencies have no direct physical meaning, hence the terms ``propagating'' and ``evanescent'' must be understood in a mathematical sense here, defined by $|\beta|<1$ and $|\beta|>1$ respectively.

In the original Lifshitz paper\cite{lifshitz56}, the $\vkp$integral is split automatically into propagating and evanescent parts by substituting $p=i\kappa_0c/\omega$. Propagating contributions correspond to integrating $p$ from 1 to 0 and evanescent to an integral from $i0$ to $i\infty$, thus avoiding the problem. We notice furthermore that the issues related to $|\beta|=1$ occur for \emph{any} choice of $\epsilon(\omega)$ hence can have nothing to do with the problems with Nernst's theorem which all concern particular permittivity models. 

%%%%%%%%%%%%%%%%%%%%%%%%%%%%%%%%%%%%%%%%%%%%%%%%%%%%%%%%%%
\subsection{Continuity}

In the classical treatment by Casimir, the vacuum energy shift was found by summing over the cavity modes of the system\cite{casimir48}, a method developed further by van Kampen et alia by use of the so-called generalised argument principle\cite{vankampen68} and elaborated by Barash and Ginzburg\cite{barash84}. The normal modes of the cavity solve the characteristic equation of the set of electromagnetic boundary conditions which reduce to the equation $D_q=0$. At these frequencies $\varphi(\omega)$ would have poles which would cause trouble with continuity.

With permittivity models such as (\ref{Drude}) and (\ref{epDC}) where dissipation is included (i.e.\ $\epsilon(\omega)$ has a nonzero imaginary part), $D(\omega,\kp)=0$ has no real-frequency solutions \footnote{Sernelius has recently shown how the normal mode interpretation may still be applied \cite{sernelius06}.} except possibly $\omega=0$ since $r_q\neq 1$ everywhere. The same is the case with (\ref{epNoDC}) if an imaginary term is inserted in the denominator as in (\ref{epDC}) (otherwise $|r_q|=1$ at $\omega=\omega_0$). The real-valued permittivities as given in (\ref{epPlasma}) and (\ref{epNoDC}) cause $r_\text{TM}$ to diverge where $\epsilon\kappa_0+\kappa=0$, however, in transgression of the continuity criterion.

We conclude that the continuity of $\phi(\omega)$ is ensured for all $\omega\neq 0$ so long as $r_q^2$ is finite, continuous and $\neq 1$, sufficient criteria for which are that $\epsilon(\omega)$ satisfies criteria \ref{epsilonSymm} - \ref{epsilonCont}.

What remains is the point $\omega=0$. A priori, this is the interesting limit, since when $T$ is very small, the $\coth$ function in (\ref{F}) differs from unity only very close to zero frequency. As is well known, reflection coefficients are occasionally ill defined in the limit where $\omega$ and $c\kp$ both approach zero, as is the case for the Drude model TE reflection coefficient, for example. $r_q^2$ is always \textit{bounded}, however, so the integrand of $\phi(\omega)$ approaches zero in this limit due to the factor $\kp$ stemming from the isotropic infinitesimal $d^2\kp = 2\pi \kp d\kp$ for any $\beta\neq1$. Hence $\phi(\omega)$ is continuous for all $\omega$ if $\epsilon(\omega)$ obeys criterion \ref{epsilonCont}. 

A more serious problem is caused by the simple poles of $\coth(\omega/2\omega_T)$ and its $T$-derivative at $\omega = 0$. As argued previously, the imaginary part of $\phi(\omega)$ is zero at $\omega=0$, so the integrand of (\ref{integral2}) does not diverge, but is in some cases finite in this limit. For sufficiently small $\omega$ and finite $\sigma$, (\ref{Drude}) and (\ref{epDC}) both have the form $\epsilon\sim A+iB/\omega$ where $A$ and $B$ are constants, while if $\sigma=0$ (\ref{epDC}) instead has the form $\epsilon\sim A + iB\omega$. In both cases the imaginary part of $r_\text{TE}^2$ falls off quickly, as $\omega^3$ and $\omega^5$ respectively, but when $A\neq 1$, $\Im\mathrm{m}\{r_\text{TM}^2\}$ decreases only linearly. One easily verifies that with respect to $\omega$, $\Im\mathrm{m}\{\phi(\omega)\}\propto \Im\mathrm{m}\{r_q^2(\omega)\}$ to leading order, hence the TM mode term of $\phi(\omega)$ is proportional to $\omega$ in the above mentioned cases. 

\begin{figure}[tb]
  \includegraphics[width=3in]{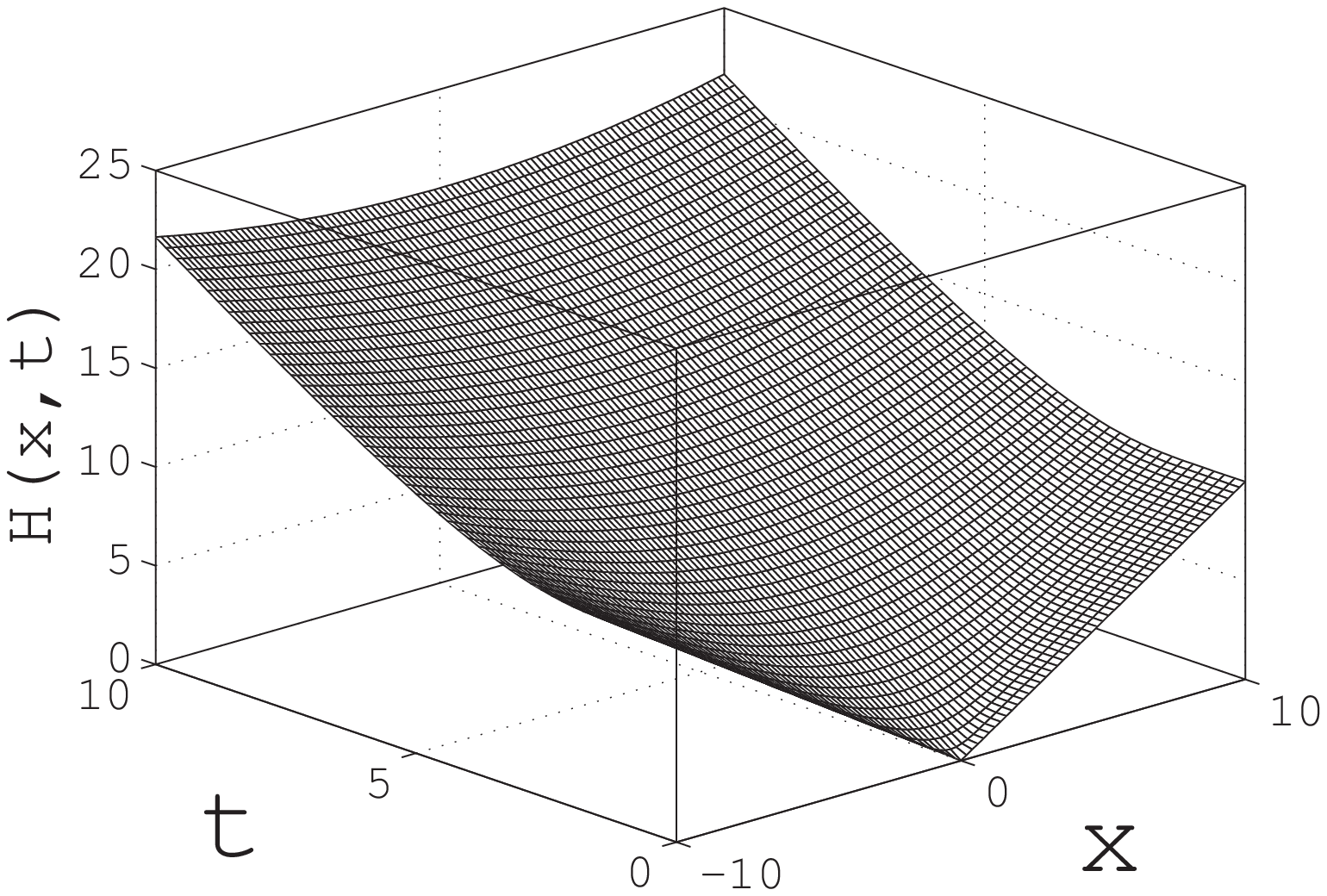}\\
  \includegraphics[width=3in]{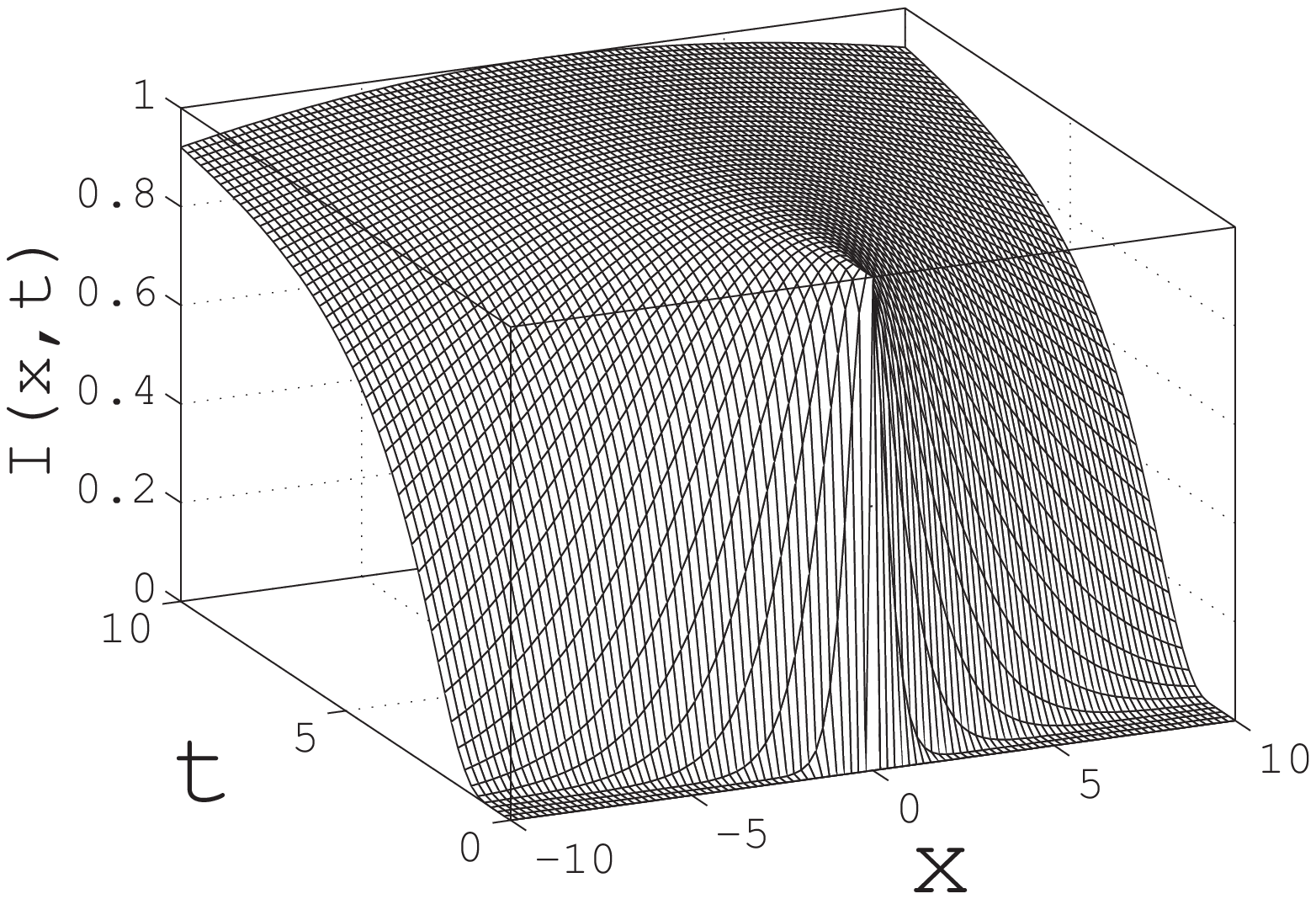}
  \caption{Plot of (\ref{H}) and (\ref{I}) as functions of $x$ and $t$. The functions $H$ and $I$ are essentially the integrands of (\ref{F}) and (\ref{integral2}) when $\phi(\omega)\propto \omega$.}\label{fig}
\end{figure}

To see how this is troublesome, consider the functions
\begin{align}
  H(x, t) &= x\frac{e^{x/2t}+e^{-x/2t}}{e^{x/2t}-e^{-x/2t}}\label{H}\\
  I(x, t) &= \frac{x^2 e^{x/t}}{t^2(e^{x/t}-1)^2},\label{I}
\end{align}
which are essentially (up to a constant factor) the integrands of (\ref{F}) and (\ref{integral2}) respectively when $\phi(\omega)\propto\omega$; here $x$ and $t$ are suitably non-dimensionalised frequency and temperature respectively so that $\omega/\omega_T=x/t$. Notice that $\partial H/\partial t = 2 I$. Equations (\ref{H}) and (\ref{I}) are plotted in figure \ref{fig} for $x$ and $t$. If we define $H(0,0)$ to be its limiting value 0, the integrand of (\ref{F}) is continuous for all $\omega$ and $T$ as we hoped, but its $T$-derivative (essentially $I(x,t)$) is not. As $T\to 0$, the integrand of (\ref{integral2}) becomes a spike of finite height and zero width. The integral past this spike is clearly zero (so the entropy would be zero as concluded above), but we run into trouble with the continuity condition. Were the limit $T\to 0$ to be taken prior to integration, $I(0,0)$ would be zero instead and $I$ would be continuous with respect to $x$ but discontinuous with respect to $t$. 

Physically it does not matter whether $I(0,0)$ is 1, 0 or something in between, since the contribution from this single point is zero in either case. Hence this one point should not matter. Formally we could state this by excluding the point $\omega=0$ from the $\omega$ integrals (\ref{F}) and (\ref{integral2}). Furthermore it is well known that the notion that \emph{every} $\epsilon(\omega)$ on the form $A+iB\omega$ or $A+iB/\omega$ would violate Nernst's theorem is incorrect; on the contrary we argue that so long as $A$ and $B$ are temperature independent, none of these will. Using an analytical software such as Maple\texttrademark\ it is quick to check that integration of $H(x,t)$ with respect to $x$ followed by differentiation with respect to $t$ gives the same result as when the order of the operations is reversed. While this argument is not rigorous it should convince the reader that the continuity issues at zero temperature and frequency can be avoided since this point is of no physical significance. 

A formally similar problem emerges when the permittivity is temperature dependent all the way down to zero temperature, as we will see, and in the latter case the singularity at zero temperature and frequencies \emph{does} appear to give a physical contribution and cannot be ignored.

\subsection{Uniform convergence}

An improper integral (\ref{leibnitz}) is said to converge uniformly (\cite{whittaker62} \S4.42) if $\forall \varepsilon>0$ there exists a number $a_0>0$ independent of $T$ such that for all $a,a'\geq a_0$,
\[
  \left| \int_a^{a'}dx f(x,T) \right| < \varepsilon.
\]
Let us briefly analyse the behaviour of $\Ln D_q(\omega,\kp)$ as $|\omega|$ and $\kp$ approach infinity. The existence of the free energy integral (\ref{F}) itself is well known, hence we need but check explicitly whether the integral (\ref{integral2}) converges uniformly along different directions in the $\omega,c\kp$ plane; clearly, if the double integral over $\omega$ and $\kp$ converges uniformly, the $\omega$ integral (\ref{integral2}) does so as well. 

As argued we consider propagating and evanescent contributions separately, in which case uniform convergence is straightforward to check. Reflection coefficients fall off rapidly as $|\omega|\to \infty$ (e.g.\ for the Drude model the real and imaginary parts of $r_q^2$ fall off as $\omega^{-4}$ and $\omega^{-5}$ respectively) and for $|\beta|>1$, $\kappa_0$ is real and positive so the integrand furthermore decreases exponentially. The factor $\exp(-2\kappa_0a)$ is oscillatory for $|\beta|<1$, but the Dirichlet integral $\int_0^\infty d x \sin x / x$ is known to be uniformly convergent, and our integrand converges more quickly than this. It is easy to check that this also holds as $|\beta| \to 0$ and $ |\beta|\to \infty$.

The splitting of (\ref{phi}) into propagating and evanescent parts may be done by integrating each part of the plane and taking the relevant limit to $c\kp\to|\omega|$ in the end. Convergence problems are then avoided for the imaginary part of $\Ln D_q$ in (\ref{integral2}); reflection coefficients fall off rapidly and further help is provided by the factor
\[
  \frac{\exp(\omega/\omega_T)}{(\exp(\omega/\omega_T)-1)^2}\approx \exp(-|\omega|/\omega_T); ~~|\omega|\gg \omega_T.
\]
The rate of convergence due to this factor depends on $T$, hence apparently cannot be used to demonstrate uniformity. We are interested only in low temperatures, however, so by defining a finite upper temperature limit $\tilde{T}$ above which the formalism is not valid, $a_0$ can be made $T$ independent (dependent on $\tilde{T}$ only). The fact that the convergence of this factor alone is not uniform for infinite temperature is unproblematic, of course.

Thus we conclude that Nernst's theorem is satisfied for $T$ independent $\epsilon(\omega)$ satisfying 1-3. The violation of Nernst's theorem in temperature dependent cases, as we shall see, can be understood as a direct consequence of violating the continuity criterion of Leibnitz' rule for improper integrals.

%%%%%%%%%%%%%%%%%%%%%%%%%%%%%%%%%%%%%%
\section{Temperature dependent permittivity}

In many models used in solid state physics, $\epsilon$ is temperature dependent for all temperatures, and herein lies the source of much of the controversy over what is the correct theory of the Casimir force between plates of real materials. The reader should note that the above theory only requires that permittivity be temperature independent for a finite temperature interval \emph{close to zero} temperature. Rather than rigorously generalising all of the above, suffice it here to discuss how the introduction of temperature dependent permittivity illuminates the entropy problems that emerge and hints at possible resolutions. In the following we will think physically in terms of positive frequencies, bearing in mind that negative frequencies exert mathematically equivalent behaviour through the symmetry criterion \ref{epsilonSymm}.

The models which have caused bother so far are the TE mode of (\ref{F}) using the Drude model (\ref{Drude}) when $\nu(T)\to 0$ as $T\to0$ (perfect lattice, no impurities) and the TM mode for dielectrics (\ref{epDC}) with $\sigma(T)\to 0$. The two different cases share many common traits, so analysing one of them in detail will suffice as illustration. Since the Drude model has already been treated in numerous efforts by both sides of the dispute (e.g.\ \cite{klimchitskaya01,brevik05}), we choose the dielectric for the below discussion.

\begin{figure}[tb]
  \includegraphics[width=3in]{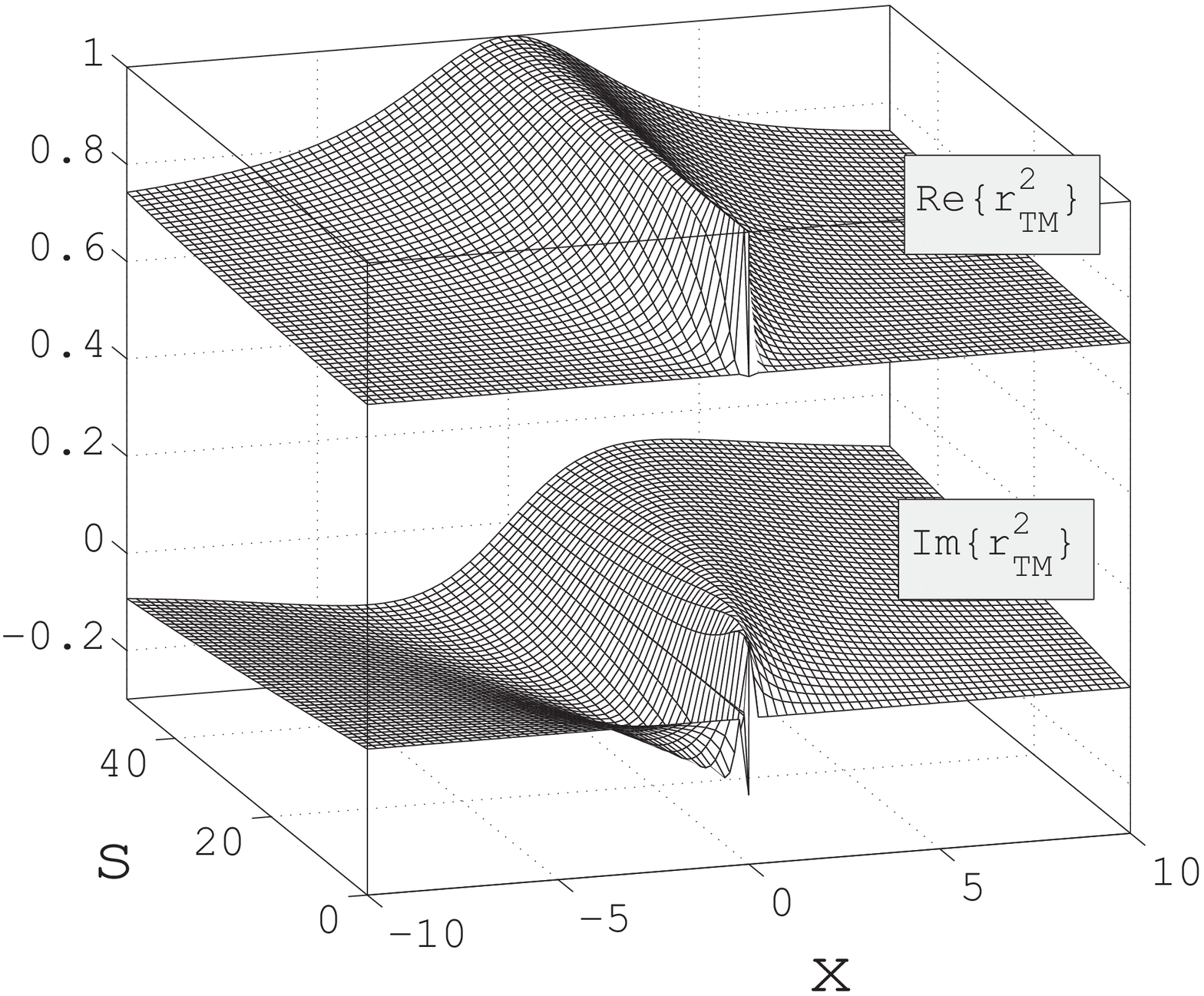}\\
  \includegraphics[width=3in]{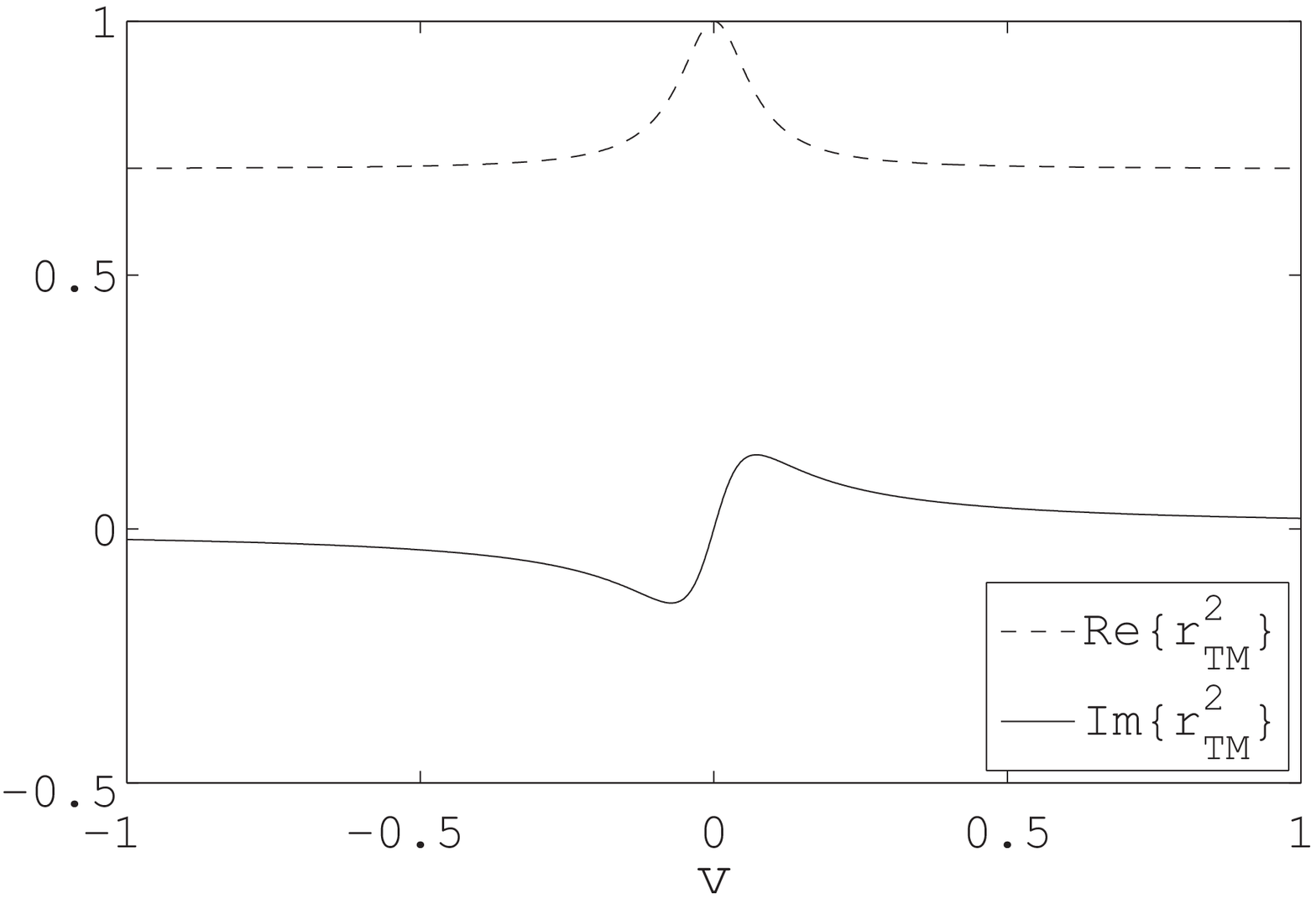}
  \caption{Plots of the real and imaginary parts of $r^2_\text{TM}$ with (\ref{epDC}) for $\omega,\sqrt{\omega\gamma}\ll \kp c, \omega_0$ and $\omega\sim \sigma/\epsilon_0$ as simplified in (\ref{epDCsimple}) with $\epsilon_\infty =11.66$. Top: as function of independent $x$ and $s$. Bottom as function of $v=x/s$.}\label{figr}
\end{figure}

As should be clear by now, the troubles with entropy emerge for small frequencies at low temperatures. Let us from now on consider the only interesting frequency range in which $\omega^2, \gamma\omega\ll \kp^2c^2,\omega_0^2$, but making no assumptions about the relative magnitude of $\omega$ and $\sigma/\epsilon_0$. Physically this corresponds to bringing $\omega$ and $T$ close enough to the limit so that for all quantities which depend on their absolute values separately they may be replaced by zero, and only quantities which depend on their relative values, specifically the TM reflection coefficient, remains in question. In this case $r_\text{TM}$ with (\ref{epDC}) inserted simplifies to
\be
  r_\text{TM} \approx  \frac{\frac{i\omega\epsilon_0}{\sigma}(\epsilon_\infty -1)-1}{\frac{i\omega\epsilon_0}{\sigma}(\epsilon_\infty +1)-1}=\frac{iv(\epsilon_\infty -1)-1}{iv(\epsilon_\infty +1)-1}\label{epDCsimple}
\ee
where $v \equiv \omega\epsilon_0/\sigma\equiv x/s$ where $x$ and $s$ are again suitably non-dimensionalised variables proportional to $\omega$ and $\sigma$ respectively. We have plotted the real and imaginary part of the squared reflection coefficients as is shown in figure \ref{figr} for illustration using $\epsilon_\infty =11.66$ as reported for Si in \cite{chen07}. 

We find that $\Re\mathrm{e}\{r^2_\text{TM}\}$ is $r_0^2\equiv(\epsilon_\infty -1)^2/(\epsilon_\infty +1)^2\approx0.71$ except for $x\ll s$ ($v\approx 0$) where it is unitary. Likewise $\Im\mathrm{m}\{r^2_\text{TM}\}$ for small $s$ is approximately zero for the most part but increases to an extremum for small $|v|$ and thence decreases linearly through $0$ at $v=0$, the same linear behaviour that led to $\Im\mathrm{m}\{\phi(\omega)\}\propto\omega$ and the discontinuity of figure \ref{fig} in the TM case before, which we argued was not essential. This time, however, there are \emph{additional} discontinuities as $s\to 0$ (equivalent to $\sigma\to 0$). In particular $\Im\mathrm{m}\{r^2_\text{TM}\}$ (which contributes to the integrals) is 0 everywhere except at $x=0$ where it can take any value between its maximum and mininum ($\approx\pm0.079$ for $\epsilon_\infty =11.66$). 

Now remember that the imaginary part of the squared reflection coefficient shown in figure \ref{figr} is to be multiplied with either the coth factor or its $T$ derivative, both of which diverge as $T/\omega$ as $\omega\to 0$. The result is an exceedingly volatile behaviour of the integrands of (\ref{F}) and (\ref{integral2}) near zero frequency and temperature, and the limit where both are zero can take any value between $-\infty$ and $\infty$ depending on the way the limit is taken! This contrasts the bounded discontinuity shown in figure \ref{fig} in the temperature independent case.

Furthermore, when $\epsilon$ contains temperature dependent parameters, entropy (\ref{integral2}) will have an additional term 
\be\label{Tterm}
  \frac{\hbar}{8\pi^2i}\infintw  \qsum \frac{e^{-2\kappa_0a}\coth \frac{\omega}{2\omega_T}}{1-r_q^2e^{-2\kappa_0a}} \cdot \frac{\partial(r_q^2)}{\partial T}
\ee
Additional entropy problems stem from this term. From (\ref{epDCsimple}) one finds with a little algebra that
\be\label{dr2dT}
  \frac{\partial}{\partial T}(r_\text{TM}^2) = -4iv\frac{iv(\epsilon_\infty -1)-1}{[iv(\epsilon_\infty +1)-1]^3} \frac{1}{\sigma}\frac{\partial \sigma}{\partial T}.
\ee
Assuming conductivity at low temperatures to behave as $\sigma_0\exp(-T_0/T)$ with $\sigma_0$ a constant (see below),
\[
  \frac{1}{\sigma}\frac{\partial \sigma}{\partial T} = \frac{T_0}{T^2}.
\]
When $\omega\neq 0$ this inverse quadratic temperature dependence is no problem since (\ref{dr2dT}) varies as $v^{-1}\propto\sigma$, so the term (\ref{Tterm}) is zero by a good measure when $T=0$. The limit $\omega\to 0$ may be taken so that $v$ has a finite value, however, in which case the derivative (\ref{dr2dT}) diverges as $T^{-2}$. This corresponds to $r_\text{TM}^2$ making a sudden jump from $1$ to $r_0^2$ at $T=0$ for $\omega=0$. The term $\partial (r^2_q)/\partial T$ at $T=0$ is thus zero for all frequencies except $\omega=0$, where it is infinite. We can no longer argue that this one point does not contribute to the physical quantity $S$, and while the purported zero temperature entropy may be difficult to calculate in this way (it is straightforward to calculate it in the imaginary frequency framework), it seems likely that entropy obtains a finite value assuming the formalism may be taken at face value. 

\subsection{Findings of the mathematical analysis}

An important realisation is thus that a violation of the third law of thermodynamics is predicted in the present framework when $\epsilon(\omega)$ changes the power of its leading order term with respect to $\omega$ at \emph{exactly} zero temperature. When $\epsilon(\omega)$ changes from $\propto\omega^{-2}$ to $\propto\omega^{-1}$, a violation occurs in the TE mode; when the change is from $\propto\omega^{-1}$ to $\propto\omega^0$, the TM mode gives the zero temperature entropy contribution.

The bottom line is that when viewed in the framework of real frequencies, all apparent zero temperature entropy anomalies stem from divergencies of the Lifshitz integrand in the double limit $T\to 0$ and $\omega \to 0$. 
%Physically, this is reasonable: It is well known that the entropy problems arise because the transition from finite to zero values of $\nu$ and $\sigma$ in the Lifshitz theory is not smooth, hence nonzero entropy at zero temperature is predicted in the special case where this discontinuous behaviour occurs at exactly zero temperature. 
The reader should note that in this author's understanding, this result does not contradict the findings of either \cite{bostrom00,hoye03,bostrom04, brevik05,brevik04,hoye07} or \cite{klimchitskaya01,geyer05, geyer08}. The latter of these references also notes a formal violation of Nernst's theorem due to rotation of permanent ionic dipole moments in the materials. In light of the above we may conclude that the latter effect is only problematic in media where the rotational degree of freedom of the ions vanishes with temperature in a manner so that its resonant frequency is finite at $T>0$ and zero at $T=0$. This anomaly needs to be studied further in the future.

%Let us assume the only temperature dependent parameters are $\nu$ in (\ref{Drude}) and $\sigma$ in (\ref{epDC}) which for perfect lattice metals and true dielectrics vanish as $T^5$ (Bloch-Gr\"uneison) \cite{hoye03} and as $\exp(-T_0/T)$ \cite{geyer05} respectively, $T_0$ being for intrinsic semiconductors half the band gap over $\kB$ (for Mott-Hubbard dielectrics it has a more complicated interpretation), in the order of magnitude of 6000K for semiconductors. A highly useful overview is provided in the forthcoming \cite{geyer08}. What one finds at $T\to 0$ is that when this behaviour is assumed, $\lim_{\omega\to 0} r$ is ill defined for the TE mode in the Drude case for $T\to 0$, and the same for the TM mode in the case of a dielectric. Specifically, $r_\rmTE^2=1$ in the $\omega,T\to 0$ limit if $T\to 0$ is taken first, but zero the other way around, likewise $r_\rmTM^2$ for dielectrics is either $(\epsilon_\infty -1)^2/(\epsilon_\infty +1)^2$ or $1$, depending on the way the limit is taken. The exact behaviour in this limit is highly non-trivial, yet we note that mathematically there is reason to doubt whether the criterion that the integral be continuous for all $T$ \textit{including} $T=0$ is satisfied. It seems the method as presented is not well suited to determine these cases, which have been argued by others to violate the Nernst theorem.

%%%%%%%%%%%%%%%%%%%%%%%%%%%%%%%%%%
\subsection{Physical discussion of thermodynamical anomalies}

In this section we will undertake a brief physical discussion of the mathematical results in the previous sections, reviewing the temperature debate for metals and semiconductors in light of the above analysis. %We argue first that there is a need to distinguish between idealisations and approximations and use this notion to discuss the physical interpretations which result when taking the the Drude-like permittivities of (\ref{Drude}) and (\ref{epDC}) and extending them to zero temperature as we have above. Thereafter we consider the case of semiconductors and argue why a solution in between the alternatives (\ref{epDC}) and (\ref{epNoDC}) is probably required, possibly through the consideration of spatial dispersion.

%The notion that $\nu$ and $\sigma$ in the Drude model for metals and (weakly conducting) semiconductors respectively vanish in the absence of thermal excitations stem from standard models found in most solid state physics textbooks. 

Models used when studying the physics of real systems are founded on assumptions which we may categorise as modelling idealisations and approximations in the description of the behaviour of these models, and there may be a need to distinguish between these in the present context. A relevant modelling idealisation, for example, is that a metal sample has infinitely large dimensions and a perfect crystal lattice structure. Much of science is founded on such ideal models and corrections to them. A relevant approximation in this context is the use of simple dielectric functions such (\ref{Drude}) to (\ref{epNoDC}) which in particular assume that the media in question have local dielectric response (i.e.\ they only depend on frequency, not momentum $\mathbf{k}$). Even for idealised systems, such approximations typically have limitations.% (see e.g.\ \cite{agranovich66}). 

An ideal model which can in principle be realised (notwithstanding infeasibility in practice) cannot be allowed to violate the laws of nature, thermodynamics in particular. An approximation, on the other hand, will typically have a finite range of applicability, and cannot be expected to behave correctly outside this range. Given that the limits $T=0$ and $\omega=0$ are in some ways extreme cases, it is especially important to investigate the latter point in relation to the purported problems with the third law of thermodynamics. 

Specifically, if an approximation which works well at room temperature does not hold for $T=0$ one cannot conclude from a formal violation of Nernst's theorem that it cannot be used \emph{within} its applicability range.

%Nernst's theorem only concerns the system's behaviour at $T=0$, however, which could be very different from that found at room temperature. If a model may be shown to violate thermodynamics at $T=0$, it is therefore important to verify that the model is applicable in this special case. If criteria for the use of the model do not hold here, zero temperature is a situation outside the model's range of applicability, and Nernst's theorem is no longer a useful test of the model's correctness, since no theory can be expected consist with the laws of nature outside of its applicability range. %Correspondingly, a theory cannot be rejected because it fails in cases where it is inapplicable.

%We have shown that the zero-temperature entropy reported in \cite{klimchitskaya01} and \cite{geyer05} may be associated with the notion that $\nu$ and $\sigma$ are temperature variant all the way down to exactly zero temperature. If on the other hand $\nu$ and $\sigma$ reach some constant value, \emph{any} value, either finite or zero, at some \emph{finite} temperature, however small, divergencies vanish as we have demonstrated and the Nernst theorem is not violated. A third option is that the local form of the dielectric response, while a good description at room temperature, is not applicable at very low temperatures.

The much investigated temperature anomaly for metals is a good example of the above, and we will review it briefly for illustration. For a perfect and infinitely large metal lattice, relaxation $\nu(T)$ is due to electron-phonon interactions only and follows the Bloch-Gr\"{u}neisen formula (see appendix D of \cite{hoye03}), according to which $\nu$ vanishes as $T^5$ as temperature tends to zero leading to the above reported anomalies. It has been pointed out that no real metal sample is ever perfect \cite{hoye03} nor infinitely large\footnote{The conductivity of very pure metals at low temperatures is found experimentally to be sample size dependent\cite{burns85} so even assuming perfectly pure metal, $\nu$ still reaches a finite value when its Bloch-Gr\"{u}neisen mean free path becomes comparable to sample dimensions. }, so relaxation does not vanish in real systems. There is now consensus that for impure metals the Drude relation does not lead to thermodynamic inconsistencies. 

However, the theoretical problem is thereby only halfway solved, because as pointed out \cite{klimchitskaya01} the fulfilment of the laws of nature cannot hinge upon the presence of imperfections: the ideal system should accord with thermodynamics as well. The solution according to the authors of \cite{klimchitskaya01} is to remove the relaxation from the Drude model and, more recently, introduce dissipation instead through a generalised plasma model \cite{geyer07}, unfortunately at the expense of ignoring the manifest presence of relaxatation at room temperature. Experiments seem to confirm the predictions from such a procedure and rule out those implied by the use of the Drude model (\ref{Drude}) (e.g.\ \cite{decca07}) but the theory has not been universally accepted.

While the ideal crystal lattice, when treated in all detail, should certainly be found to abide by Nernst's theorem, the approximation that its dielectric response is well described by the local formalism has been questioned for temperatures approaching $T=0$. Svetovoy and Esquivel \cite{svetovoy05} and Sernelius \cite{sernelius05} conclude independently that at low temperatures nonlocal effects (the anomalous skin effect) dominate, and the local models are no longer reliable. Their analyses accounting for spatially dispersive effects reveal that, within the approximations made in \cite{svetovoy05, sernelius05} Nernst's theorem is fulfilled independently of the presence of imperfections as it should. The spatial dispersion approach was criticised \cite{klimchitskaya05} on several accounts with reference to a treatment by Barash and Ginzburg many years ago\cite{barash75} (see also \cite{agranovich66}). The paradox remains that such a careful procedure (albeit not free of approximations) does not accord with available experimental data. Commendable efforts at a resolution include the recent exploits by Bimonte \cite{bimonte07}.

%to calculate $\nu(T)$ one finds for gold at 0.5K, for example $\nu\approx 3\cdot 10^{3}$ rad/s, corresponding to a mean free path for conduction electrons at Fermi energy (velocity $\approx 1.4\cdot 10^6$m/s \cite{burns85}) of around 3 kilometres. Even when the metal lattice is otherwise perfectly free of defects and impurities, electrons will be scattered by the edges of the metal plate causing relaxation. Indeed %At very low temperature, thus `infinite plates' no longer simply means 'macroscopic' as is usually the case: Assuming relaxation follows Bloch-Gr\"{u}neisen, infinite size might be a good approximation when electronic relaxation times are comparable to the duration of an experiment (at $T=0$), say one second. The corresponding mean free path would be $1,400$ kilometres in gold, roughly the distance from London to Rome.

%The problem with the latter argument, as pointed out before \cite{klimchitskaya01}, is that the fulfilment of the laws of nature cannot hinge upon the presence of imperfections such as lattice defects or finite boundaries. 

While mathematically analoguous to the metal case, the temperature anomaly for semiconductors is physically different. Here the problem is not related to idealised models (the conductivity of isolators truly does vanish at zero $T$), but approximations only. 

One can argue intuitively that the approximate model (\ref{epDC}) probably cannot be taken at face value when conductivity is very small since it implies that even an infinitesimal conductivity should give rise to large thermal corrections in the Lifshitz formalism contrary to physical intuition. For isolators, by definition, conductivity vanishes at $T=0$, and for many such materials conductivity even at room temperature is so small it would be expected to make for a minor perturbation only \footnote{For an introduction to different types of semiconductors, see chapter 1 of \cite{gebhard97}}. If (\ref{epDC}) is a poor approximation at low frequencies as $\sigma$ vanishes, its violating thermodynamic laws in this case may not be too worrisome. %In a slab of finite size there will at some low temperature be too few mobile electrons to perfect the screening of an electrostatic field, and $r_\text{TM}^2(\zeta=0)$ should drop below the value $1$ implied by (\ref{epDC}). 

A recent experiment \cite{chen07} measured the force between a substrate of the semiconductor silicon and a gold sphere. The semiconductor was excited into a metallic state by a pulsed laser and it was concluded that while the model (\ref{epDC}) was a good representation in the metallic state, the inclusion of the $\sigma$ term when the material was in the poorly conducting state was excluded at 95\% confidence. This conclusion may not be surprising in light of the above argument which indicates that a Drude type permittivity model overestimates the effect of a small conductivity in the Lifshitz formalism. If so, it is likely that the experimental result might be explained without reference to the Nernst theorem which concerns physics far removed from laboratory conditions.

Attempts at a more careful description of the effect of a small density of free charges were recently made by Pitaevskii \cite{pitaevskii08} and by Dalvit and Lamoreaux \cite{dalvit08}, based on the effects of Debye-H\"uckel screening from free charges in mean field theory. The resulting expressions do not fit experimental data well \cite{klimchitskaya08}, and it is possible that a more detailed screening model is needed. %a possible explanation for which could be that the modelled response of the free electrons, based on the effects of Debye-H\"uckel screening from free charges, assumes that the free electrons are able to rearrange themselves to equilibrium configuration. If mobility is limited, electromagnetic out-of-equilibrium effects could diminish the screening effect.

\section{Conclusion}

While the Lifshitz formalism at real frequencies is much more complicated than the imaginary frequency equivalent normally considered, the consideration of quantities with direct physical interpretations sheds new light on the problem of non-vanishing Lifshitz entropy at zero temperature. We have argued that Nernst's heat theorem is not violated for any causal and continuous (except at $\omega=0$) $\epsilon(\omega)$ which is independent of $T$ near $T=0$. This accords with the findings of Intravaia and Henkel\cite{intravaia07} using a different approach. More generally, this holds for dielectric plates whose squared Fresnel TE and TM reflection coefficients are continuous for all $\omega$ and $T$ and non-unitary except at $\omega=0$. It follows from this that entropy anomalies previously reported pertain to the persistence of the permittivity's temperature variance all the way to zero temperature and are consequences of divergencies in the Lifshitz formalism in the double limit $\omega\to 0$ and $T\to 0$. 

When considering physical consistency in such limits as a means to distinguish between candidate theories, particular care must be taken. We emphasise that approximations can only be judged based on their performance within their domain of applicability. Specifically, approximations which are invalid at $T=0$ cannot be expected to be well behaved in this limit, hence cannot be rejected for causing violation Nernst's theorem which concerns zero temperature only. It is therefore important to verify carefully that approximate physical models probed by invoking Nernst's theorem are valid in this case. We finally argue that a recent experiment using an optically excited semiconductor can probably be explained without reference to the Nernst theorem by accounting for the presence of free charges more carefully than by the use of local Drude-type models.

\section*{Acknowledgements}

The author thanks Professors Iver Brevik, Kimball Milton and Dr.\ Francesco Intravaia for discussions and useful comments. Further thanks to Professor Vladimir Mostepanenko for many useful remarks on the manuscript and for kindly allowing a preview of \cite{geyer08}.

\end{document}